# Optimal Efficiency of Self-Assembling Light-Harvesting Arrays[†]

## Ji-Hyun Kim and Jianshu Cao*

*Department of Chemistry, Massachusetts Institute of Technology, Cambridge, Massachusetts 02139, United States*

*Received: July 22, 2010; Revised Manuscript Received: September 29, 2010*

Using a classical master equation that describes energy transfer over a given lattice, we explore how energy transfer efficiency along with the photon capturing ability depends on network connectivity, on transfer rates, and on volume fractions—the numbers and relative ratio of fluorescence chromophore components, e.g., donor (D), acceptor (A), and bridge (B) chromophores. For a one-dimensional AD array, the exact analytical expression (derived in Appendix A) for efficiency shows a steep increase with a D-to-A transfer rate when a spontaneous decay is sufficiently slow. This result implies that the introduction of B chromophores can be a useful method for improving efficiency for a two-component AD system with inefficient D-to-A transfer and slow spontaneous decay. Analysis of this one-dimensional system can be extended to higher-dimensional systems with chromophores arranged in structures such as a helical or stacked-disk rod, which models the self-assembling monomers of the tobacco mosaic virus coat protein. For the stacked-disk rod, we observe the following: (1) With spacings between sites fixed, a staggered conformation is more efficient than an eclipsed conformation. (2) For a given ratio of A and D chromophores, the uniform distribution of acceptors that minimizes the mean first passage time to acceptors is a key point to designing the optimal network for a donor−acceptor system with a relatively small D-to-A transfer rate. (3) For a three-component ABD system with a large B-to-A transfer rate, a key design strategy is to increase the number of the pathways in accordance with the directional energy flow from D to B to A chromophores. These conclusions are consistent with the experimental findings reported by Francis, Fleming, and their co-workers and suggest that synthetic architectures of self-assembling supermolecules and the distributions of AD or ABD chromophore components can be optimized for efficient light-harvesting energy transfer.

## I. Introduction

Photosynthesis, an essential ecological process, efficiently converts light energy into chemical energy through various membrane complexes. This conversion process has inspired many researchers in developing light-harvesting devices.[1] The high performance of efficiency for photosynthetic networks found in nature can be related to the spatial distribution of various chromophores that transport energy to reaction centers via a series of fluorescence resonance energy transfers. There have been many studies focusing on such aspects for various photosynthetic networks found in nature.[2–5] To capture photons efficiently, antenna sites should occupy a large area. However, too many antenna sites for each reaction center result in a decrease in efficiency because it takes long times for excitations to find the reaction centers so that the chance of excitation degradation through irrelevant channels increases. This observation reveals that an optimal efficiency can be established by varying the donor−acceptor ratio.

Inspired by these naturally occurring energy transfer networks, much effort has been devoted to designing artificial light-harvesting architectures.[6–9] Such assemblies are both promising for applications to devices such as solar cells, and for identifying the controlling factors for optimal networks. A new synthetic method to assemble an artificial architecture has been developed using the self-assembling property of the tobacco mosaic virus coat protein monomers.[10,11] The helix or disk structures synthesized in these experiments have donor−acceptor systems that show strong dependency on the number ratio of donors (D) to acceptors (A). Furthermore, the incorporation of bridge chromophores (B), which is similar to a doping mechanism for semiconductors, facilitates energy transfer toward acceptors and results in the remarkable improvement in efficiency. In this paper, we explore optimal efficiency in terms of network connectivity, transfer rates related to the types of fluorescence chromophores, and numbers and relative ratio of the chromophore components. In particular, the stacked-disk rod structure found in ref 10 has a high tunability because multiple disks of like size and differing compositions can be combined. An example of a three-component tunable system can be found in light emitting devices containing a mixed-monolayer consisting of red, green, and blue emitting colloidal quantum dots, where the emission spectrum can be tuned by changing the ratio of differently colored quantum dots without changing the structure.[12] By examining several spatial arrangements of the chromophores, we can correlate structure to efficiency.

## II. One-Dimensional Systems

We begin with a one-dimensional array on a uniform lattice. For this system, since energy transfers between nonadjacent sites have no significant effects on the efficiency, we only consider nearest-neighbor hopping for simplicity. The easiest way to construct an optimal array is to compare different lengths of donor segments bounded by two acceptors such as a segment shown in Figure 1a. The corresponding classical master equation is given by

$$\frac{\partial}{\partial t}\mathbf{P}(t) = -k_S \mathbf{P}(t) - k_{DD}\mathbf{A}\cdot\mathbf{P}(t) - k_{DA}\mathbf{R}\cdot\mathbf{P}(t) \quad (1)$$

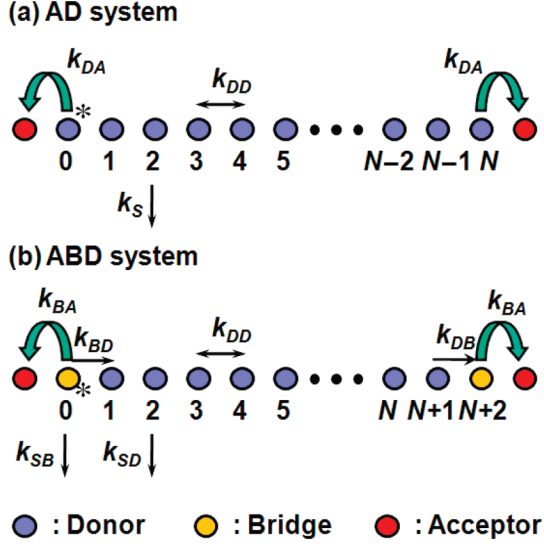

**Figure 1.** One-dimensional energy transfer systems. (a) The acceptor−donor (AD) system consisting of $N + 1$ donors and two acceptors at both ends. (b) The acceptor-bridge-donor (ABD) system has two bridge sites added to adjacent positions of two acceptors.

$\mathbf{P}(t)$ is the $(N + 1)$-dimensional column vector where the $i$th element ($0 \le i \le N$) is the probability that an excitation is located at the $i$th donor site at time $t$. Note that $\mathbf{P}(t)$ is assigned only to donors, which implies an irreversible transfer from adjacent donors to acceptors. $k_S$ is the rate constant of spontaneous relaxation. $k_{DD}$ and $k_{DA}$ denote the D-to-D and D-to-A rate constants, respectively. $\mathbf{A}$ is the $(N + 1) \times (N + 1)$ Rouse matrix defined as $(\mathbf{A})_{ij} = 2\delta_{ij} - \delta_{i-1,j} - \delta_{i+1,j} - \delta_{i,0}\delta_{0,j} - \delta_{i,N}\delta_{N,j}$. $\mathbf{R}$ is the $(N + 1) \times (N + 1)$ reaction matrix with the elements given by $(\mathbf{R})_{ij} = \delta_{i0}\delta_{0j} + \delta_{iN}\delta_{Nj}$. Integrating both sides of eq 1 over the time and multiplying the resultant by the $(N + 1)$-dimensional row vector $\mathbf{u}^T = (1, 1, ..., 1)$ from the left, one can obtain the branching relation

$$P_S + P_A = 1 \quad (2)$$

where $P_S = k_S \sum_{n=0}^{N} \int_0^\infty dt\, p_n(t)$ and $P_A = k_{DA} \int_0^\infty dt\, [p_0(t) + p_N(t)]$. $P_S$ and $P_A$ are the total probabilities for the excitation energy to decay before reaching acceptors and to reach acceptors, respectively. Using eq 2, the quantum yield $P_A$ can be simply expressed as

$$P_A = 1 - k_S \hat{S}(s=0) \quad (3)$$

where $\hat{S}(s)(= \int_0^\infty dt\, e^{-st} S(t))$ denotes the Laplace transform of the survival probability $S(t)$ ($= \sum_{n=0}^{N} p_n(t)$) that an excitation survives on the array by time $t$. We can then write our definition of light-harvesting efficiency as

$$q = P_A X_D \quad (4)$$

where $X_D$ is the fraction of donors or, explicitly, $X_D = N_D/(N_D + N_A)$. The numbers of donors and acceptors, $N_D$ and $N_A$, are $N + 1$ and 2, respectively. $X_D$ is a measure of the system's ability to capture initially injected photons. Here, we assume absorption cross sections of donors are identical and are independent of incident photon frequency. The definition of overall efficiency, eq 4, has a meaning similar to that used in the two-dimensional membrane system.[13] In ref 13, instead of $P_A$, $\eta$ was used to represent the quantum yield. $P_A$ monotonically decreases with $N_D$ while $X_D$ shows the opposite behavior, suggesting the existence of an optimal $q$.

We now consider the one-dimensional array depicted in Figure 1b. For a given pair of D and A, we expect that inserting another species (B) will increase the efficiency when the spectral overlap between B and A is larger than the spectral overlap between D and A. Since the back transfer from B to D is inefficient,[10] the design shown in Figure 1b is optimal when B is inserted to the AD system in Figure 1a. Considering the possibility that B is partially bright at the maximum absorption wavelength of the donor, we will examine the cases for when B is fully bright and for when B is fully dark. The fractions of bright species $X_{Bright}$ corresponding to the respective cases are thus $X_{D,B} = (N + 3)/(N + 5)$ and $X_D = (N + 1)/(N + 5)$. The derivations for the explicit expressions of $P_A$ for these two one-dimensional systems are given in Appendices A and C.

In ref 14, $q$ is used indistinguishably from $P_A$ because only one acceptor is considered and $N_D$ is fixed. Because $k_S$ is small compared to the other rates, the small-$k_S$ approximation of eq 3 can be obtained by setting $k_S$ in $\hat{S}(0)$ to be zero and using a [0/1]-Padé approximation:

$$P_A = \frac{1}{1 + k_S \hat{S}(0)_{k_S=0}} \quad (5)$$

where $\hat{S}(0)_{k_S=0}$ means the $k_S$-independent mean first passage time of excitation to acceptors, $\langle t_f \rangle_{k_S=0}$. $k_S$ in eq 5 corresponds to $k_d$ in the equivalent formula in ref 14. The first passage statistics of a complex kinetic scheme has been recently treated with the two equivalent formalisms, i.e., the rate matrix formalism and the waiting time distribution formalism.[15] Both formalisms can be used to calculate the mean first passage time and quantum yield. Here, we adopt the rate matrix formalism.

Figure 2a shows the existence of an optimal $N_D$, which is around 10 with the typical choice of $k_S = 10^{-2}k_{DD}$. Values of the rescaled parameters throughout this paper have been chosen on the basis of ref 11, where $k_S/k_{DD} = 0.016$ and $k_{DA}/k_{DD} = 0.37$. In Appendix A1, we present an explicit analytical expression of $\hat{S}(s)$ for calculating $P_A$ in eq A11 and a large-$N$ approximation of $q$ in eq A12. The performance of eq A12 is surprisingly good as compared to the exact results obtained using eq A11, despite a slight underestimation for large $k_{DA}$ values. As inferred from eq A12, the profile of efficiency converges into a single curve in the limit of large $k_{DA}$. Parts c and d of Figure 2 are contour plots of the optimal number of donors ($N_D^*$) and the corresponding optimal efficiency ($q^* = q(N_D^*)$) on the parametric plane ($k_S$, $k_{DA}$), respectively. Either a decrease in $k_S$ or an increase in $k_{DA}$ results in an increase both in the optimal number of donors and in the optimal efficiency. Such qualitative features can be analytically expressed using eqs A14 and A12. The increase of $q$ with $k_{DA}$ is strongly modulated by the magnitude of $k_S$. The smaller the $k_S$, the steeper the increase in the optimal number or in the optimal efficiency with $k_{DA}$. This qualitative feature clearly appears in eq 5 with noting that $\langle t_f \rangle_{k_S=0}$ decreases with $k_{DA}$.

Numerical results obtained using eq C6 for the ABD system with $k_{DB} = k_{DD}$ and $k_{SB} = k_{SD} = 10^{-2}k_{DD}$ are shown in Figure 2b. The open and filled black symbols correspond to the fully bright B and fully dark B cases, respectively. The circles stand for $(k_{BA}/k_{DD}, k_{BD}/k_{DD}) = (10, 0)$, the squares for $(1, 0)$, and the diamonds for (1, 1). The result for (10, 1) is omitted because it is nearly indistinguishable from the result for (10, 0), whereas the effect of back transfer from B to D on $q$ is noticeable for relatively small value of $k_{BA}$. The profiles for the ABD system are compared with that for the AD system, which is the curve of Figure 2a with $k_{DA}/k_{DD} = 10^{-1}$. Regardless of the brightness of B, the ABD system is always more efficient than the AD system except in the small $N_D$ regime. If a larger $k_S$ value is used, the relative increment in $q$ due to the incorporation of B decreases.

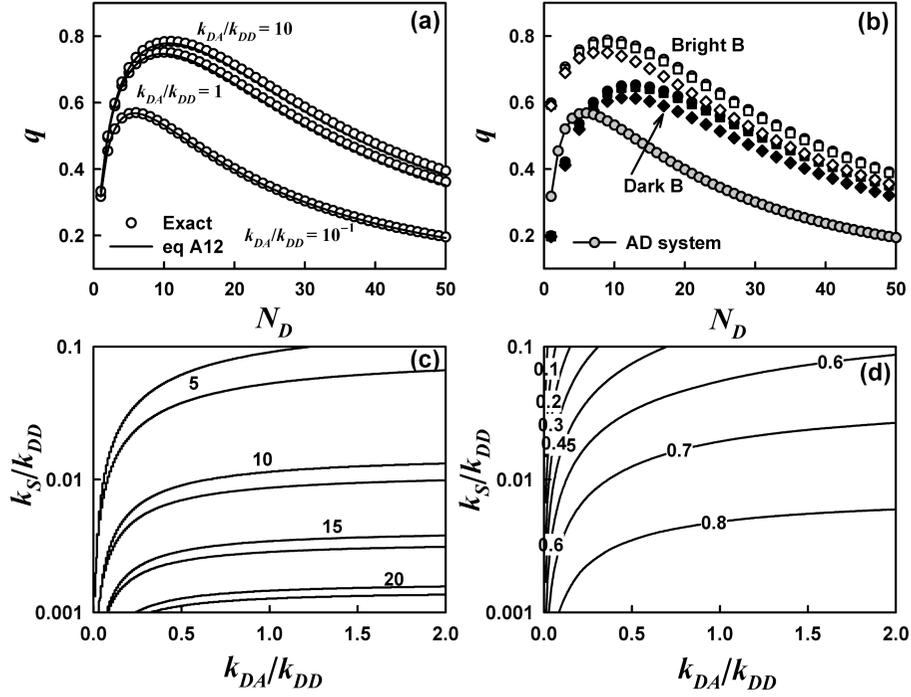

**Figure 2.** (a) Efficiency $q$ as a function of the number of donors $N_D$ for the one-dimensional AD system with $k_S/k_{DD} = 10^{-2}$. The exact results are compared to the large $N_D$-limit expression $q_\infty$ given in eq A12. (b) Efficiency $q$ as a function of $N_D$ for the one-dimensional ABD system with $k_S/k_{DD} = 10^{-2}$. The open and filled symbols represent the bright B and dark B cases, respectively. The circles, squares, and diamonds represent the parameter sets of $(k_{BA}/k_{DD}, k_{BD}/k_{DD}) = (10, 0), (1, 0)$, and $(1, 1)$, respectively. The filled gray circles represent the values of $q$ with $k_{DA}/k_{DD} = 10^{-1}$ given in Figure 2a. (c), (d) Contour plots for the one-dimensional AD system show the optimal number of donors and the corresponding optimal efficiency on the reduced parameter plane $(k_{DA}/k_{DD}, k_S/k_{DD})$, respectively. In Figure 1c, any two closely positioned curves are assigned to have the same value.

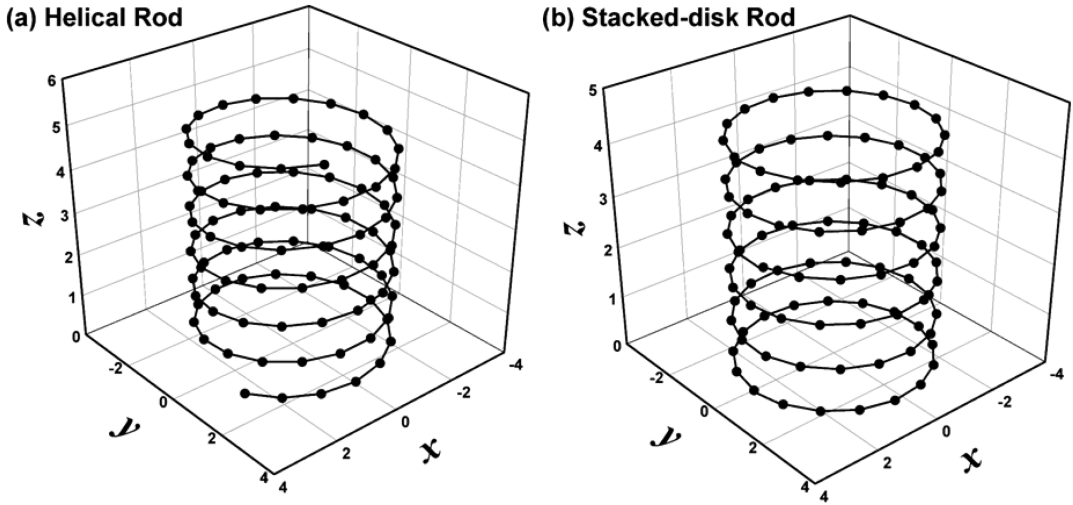

**Figure 3.** Three-dimensional rod systems including 102 sites. (a) The helical rod has the parameters, $(p, R_h) = (3^{1/2}/2, 16.5)$. (b) The stacked-disk rod has the parameters, $(h, R_d, \phi) = (3^{1/2}/2, 17, \pi/17)$.

### III. Three-Dimensional AD Rods

The one-dimensional system can be assembled into higher-dimensional systems by structural transformations such as folding, cyclization, and stacking. Folding can produce a helix that is a regular three-dimensional structure with a simple geometrical parametrization. The end-to-end cyclization of the one-dimensional array given in Figure 1a, with only the nearest-neighbor hopping considered, results in a ring system equivalent to the linear system in the sense that both systems share the same master equation, eq 1. Stacking multiple rings can also produce a three-dimensional rod system. Figure 3 shows the two rod structures based on the experiments in refs 10 and 11. For these high-dimensional systems, the effect of long-range energy transfers between nonadjacent pairs can be significant. We therefore consider the more general form of the master equation given by

$$\frac{\delta}{\delta t}p_i(t) = -\sum_j K_{ij}p_j(t) \quad (6)$$

where $p_i(t)$ is the probability that an excitation is located at the $i$th site. Note that the column vector $\mathbf{p}(t)$ has site indexes only for donors and/or bridges. The rate matrix element $K_{ij}$ is defined as

$$K_{ij} = -T_{ij} + \delta_{ij}(k_S + \sum_k T_{kj}) \quad (7)$$

where the distance-dependent transfer rate from the $j$th to $i$th sites $T_{ij}$ is defined by

$$T_{ij} = k_{mn}r_{ij}^{-6} \quad (m, n \in \{A, B, D\}) \quad (8)$$

In eq 8, $r_{ij}$ is the shortest distance between the $(i, j)$-pair, and $m$ and $n$ denote the species that the $j$th and $i$th sites belong to, respectively. For convenience, $k_{SD}$ and $k_{SB}$ are set equal to each other so that a single symbol $k_S$ is used in eq 7. The matrix element $T_{ij}$ can be easily calculated given a set of site coordinates. Henceforth, we will use the unit system where the time scale is given by $k_{DD}^{-1}$ and the length scale is given by the spacing between two horizontal nearest neighbors. The helical rod has the coordinates described as $\{r \cos \theta_k, r \sin \theta_k, (k-1)d\}$, where $d = p/R_h$ with $p$ denoting the pitch, $r = (1 - d^2)^{1/2}(2 \sin(\pi/R_h))^{-1}$, and $\theta_k = 2\pi k/R_h$. $R_h$ defines the number of sites per turn. The stacked-disk rod can be generated using $\{r \cos \theta_k, r \sin \theta_k, mh\}$, where $r = (2 \sin(\pi/R_d))^{-1}$ and $\theta_k = 2\pi k/R_d - (1 - (-1)^m)\phi/2$. $R_d$ denotes the number of sites on a disk, $m$ the quotient obtained by dividing $k - 1$ ($\geq 0$) by $R_d$, and $h$ and $\phi$ are the vertical distance and staggering angle between two adjacent disks, respectively. From eq 6, $q$ can be obtained as

$$q = [1 - k_S \sum_{i,j} [\mathbf{K}^{-1}]_{ij} p_j(0)] X_{Bright} \quad (9)$$

For AD systems, $p_i(0) = N_D^{-1}$. For ABD systems, $p_{i \in D \text{ or } B}(0) = (N_D + N_B)^{-1}$ for bright B and $p_i(0) = N_D^{-1}(1 - \delta_{i, \forall j \in B})$ for dark B. Forthcoming numerical results are calculated using eq 9.

Parts a and b of Figure 4 show the numerical results for a helical rod with $(p, R_h) = (3^{1/2}/2, 16.5)$ and 102 total sites. For a given ratio of donors to acceptors, acceptors are randomly located among the 102 sites and the resulting efficiency is averaged over sufficient realizations to reach convergence. The value of $k_S$ used in Figure 4a is 0.01, and the ratio used in Figure 4b is 16:1. Figure 4a shows the existence of an optimal $q$ in terms of the number ratio and that the optimal ratio is shifted with $k_{DA}$. For $k_{DA} = 0.1$, the variation of $q$ with the number ratio qualitatively reproduces the experimental result, which uses a related definition for efficiency, given in Figure 3e of ref 10. Although more sites are used in ref 10, explicitly, 700 chromophores per 100 nm of rod length with a vertical spacing of 2.3 nm, our results are essentially invariant even when the number of sites is extended at a constant chromophore ratio. This result has some resemblance to the minimal functional unit mentioned in ref 13. In Figure 4b, we observe that the profile of $q$ with a smaller $k_S$ reaches a plateau faster. This qualitative feature can be explained by using eq 5. Although the embedded spatial dimension is 3, the qualitative behavior of $q$ with $k_S$ and $k_{DA}$ remains the same as the one-dimensional system. The stacked-disk rod with lattice spacing similar to that of helical rod shows no significant differences from the helical rod.

We investigate how efficiency depends on spatial distributions with the disk structure. Figure 4c shows that $q$ decreases with arc distance $\alpha$ between two acceptors on a 17-membered disk, where only nearest-neighbor hopping is considered. In this case, the system is just an A-to-A combination between two different lengths of segments. $P_A$ is then given by the exact expression $P_A = P_A(N_1)(N_1 + 1)/(N_D) + P_A(N_2)(N_2 + 1)/(N_D)$, where $P_A(N)$ corresponds to eq 3 with $\hat{S}(s)$ given by eq A11. The filled circles stand for $(k_S, k_{DA}) = (10^{-3}, 10^{-1})$, the open circles for $(10^{-2}, 10^{-1})$. Figure 4c implies that for a given number ratio, the arrangement where acceptors are kept furthest from each other is more favorable. This effect has the same origin as the competition effect found in diffusion-controlled reactions.[16] $k_S$ significantly decreases $q$.

In Figure 4d, the effect of the staggering angle $\phi$ is investigated for the stacked-disk rod with $R_d = 17$ and 102 total sites. The values of $k_S$ and $k_{DA}$ are 0.01 and 0.1, and the D-to-A ratio is 16:1. With spacing between sites fixed to the basic length scale, the value of $h$ varies depending on $\phi$, explicitly, $h = 1$ at $\phi = 0$ and $h = 3^{1/2}/2$ at $\phi = \pi/17$. The left and right bars correspond to the nearest-neighbor hopping and long-range transfer, respectively. The fully staggered conformation is more efficient than the fully eclipsed one. This can be explained by considering the total transfer rate toward a single acceptor embedded in the same lattice in a system consisting only of donors, explicitly, $\Sigma_{j \in D}T_{Aj}$ ($\equiv T_A$). Here, we use the maximal $T_A$ for the rod under consideration. For nearest-neighbor hopping, $T_A$ increases from 4 $k_{DA}$ ($\phi = 0$) to $6k_{DA}$ ($\phi = \pi/17$). For long-range transfer, $T_A$ increases from 4.66 $k_{DA}$ ($\phi = 0$) to 6.35 $k_{DA}$ ($\phi = \pi/17$). The increment in $T_A$ for the nearest-neighbor hopping is more pronounced compared to that for the long-range transfer, which is reflected in the increment in $q$. Note that the effect of long-range transfer is significant for the eclipsed conformation while it becomes less important for the staggered conformation.

### IV. Three-Dimensional ABD Rods

We investigate the performance of the ABD system for stacked-disk rods, particularly, the dependence of $q$ on the spatial distribution of B. We choose $N_D:N_B:N_A = 68:28:6$ as the number ratio, which is similar to the ratio used in ref 10. For this ratio, Figure 5 shows four different spatial distributions of B. The different positions for two disks consisting only of B and A are shown in Figures 5a−c. Three acceptors are randomly positioned over each AB disk. The AB disks are designed to minimize the inefficient B-to-D back transfer. Figure 5d represents the random distributions of B and A over the whole network. The corresponding numerical values of $q$ are given in Figure 6 for bright B. The circles and downward triangles correspond to $k_{BD} = 0$ and 1, respectively. Other parameters are set as $k_{BB} = 1$, $k_{DB} = 1$, and $k_{BA} = 1$. The filled gray circles represent the AD system where all the B chromophores in two AB disks are replaced by D. Both the ABD and AD systems have $k_{DA} = 0.1$ and $k_S = 0.01$. The geometrical parameters are set as $(h, R_d, \phi) = (3^{1/2}/2, 17, \pi/17)$.

First, we present the results of the AD system. The magnitude of efficiency $q$ has the relation, $q_b > q_a \approx q_d > q_c$. This order can be explained by considering how fast an excitation reaches acceptors, and the ranking can be estimated by counting the number of donor disks directly accessible to the AD disks. For b, a, and c, the numbers of donor disks are 4, 3, and 2, respectively. The descending order is in accordance with the order in $q$. The random distribution in case $d$ can be regarded as a random positioning of two AD disks, which corresponds to an intermediate case similar to case $a$. For the ABD systems, however, the order undergoes a significant change so that $q_c > q_a \approx q_b > q_d$ for $k_{BD} = 0$ and $q_c > q_a \approx q_b \approx q_d$ for $k_{BD} = 1$. The worst distribution in case $c$ for the AD system becomes the best one for the ABD system, even though the distribution in case $b$ provides the best accessibility to the AB disks. For $k_{BD} = 1$, such a reversion occurs around $k_{BA} = 0.35$. This phenomenon can be understood by the fact that the migration time of excitation after the first arrival at the AB disk for case $c$ is as short as the fast migration time cancels out the disadvantage resulting from a relatively late arrival at the AB disk. For case $c$, an acceptor has the most B nearest neighbors among the distributions given in Figure 5. In other words, $c$ has the most direct B-to-A pathways via interdisk transfers. This argument can be confirmed by controlling the migration time through $k_{BB}$. When $k_{BB}$ increases to 10, $c$ becomes slightly more efficient than $b$ by about 0.01, which is the difference in efficiency between $b$ and $c$. However, if $k_{BB}$ decreases to 0.1, the efficiency for $c$ becomes higher by 0.16. The distribution $c$ models the ABD system found in ref 10.

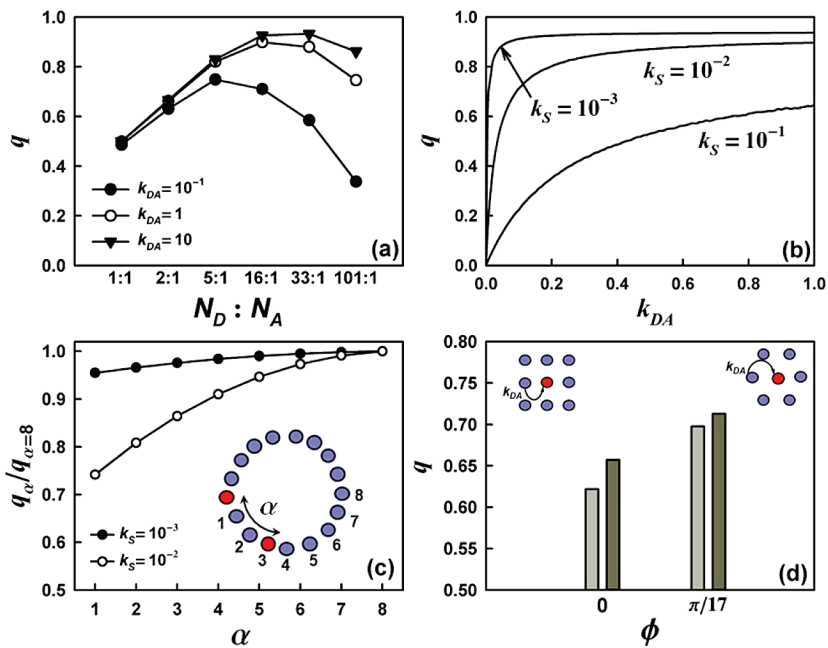

**Figure 4.** Helical AD rod system with $(p, R_h) = (3^{1/2}/2, 16.5)$ has acceptor sites that are randomly distributed over the 102 sites: The efficiency $q$ is given as a function of (a) the ratio of $N_D$ to $N_A$ and (b) $k_{DA}$, respectively. (c) Two acceptors are separated by the relative arc distance $\alpha$ on a 17-membered disk. The efficiency $q$ normalized by value at $\alpha = 8$ is shown as a function of $\alpha$. (d) Values of $q$ at the staggering angles $\phi = 0$ ($h = 1$) and $\phi = \pi/17$ ($h = 3^{1/2}/2$) are shown for the stacked-disk AD rod system with $R_d = 17$ and 102 sites. $(k_S, k_{DA}) = (10^{-2}, 10^{-1})$. The left and right bars represent the nearest-neighbor hopping and long-range transfer, respectively. The side views shown in the upper part focus on the local area around a single A for the different rods. In (b) and (d), $N_D:N_A = 16:1$.

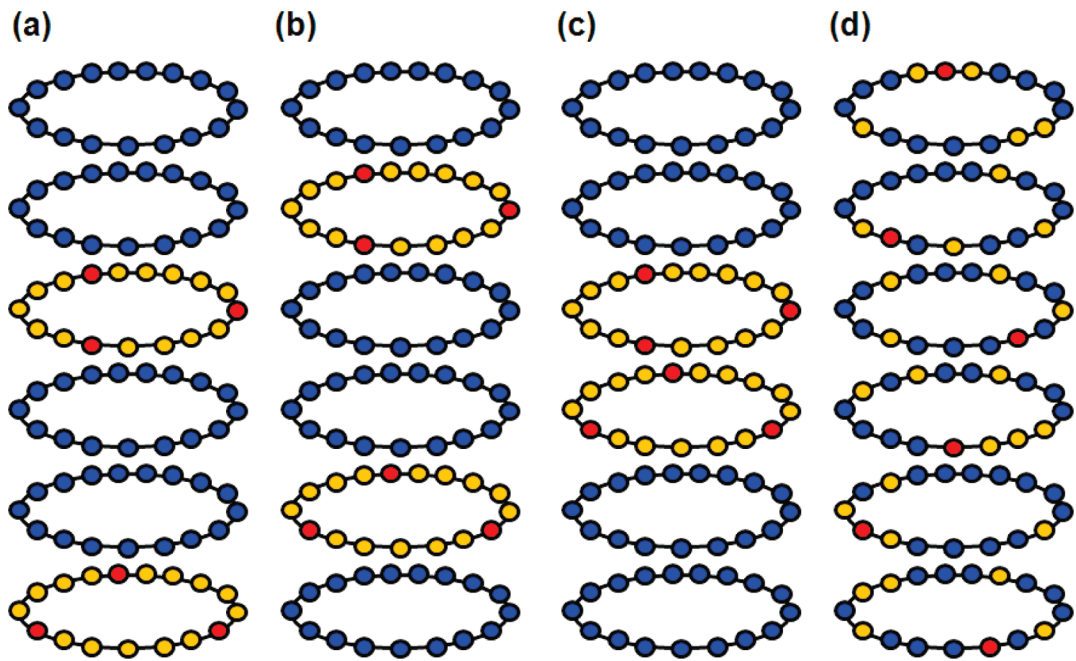

**Figure 5.** Stacked-disk ABD rod systems with the ratio $N_D:N_B:N_A = 68:28:6$. The two AB disks are positioned at (a) the first and fourth disks, (b) the second and fifth disks, and (c) the third and fourth disks, respectively. (d) The acceptors and bridges are randomly distributed over the whole sites with the same number ratio.

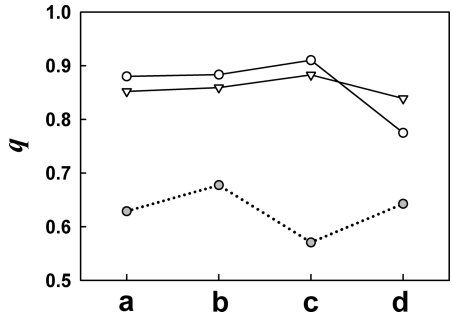

**Figure 6.** Dependence of $q$ on the spatial distribution of the respective chromophores given in Figure 5. The open symbols represent the ABD system with bright B chromophores. The circles and downward triangles represent $k_{BD} = 0$ and $k_{BD} = 1$, respectively. The remaining parameters are set as $k_{BB} = 1$, $k_{DB} = 1$, and $k_{BA} = 1$. The filled gray circles represent the AD system where all the B chromophores within two AB disks are replaced by D. Both the ABD and AD systems have $k_{DA} = 0.1$ and $k_S = 0.01$. The geometrical parameters are set as $(h, R_d, \phi) = (3^{1/2}/2, 17, \pi/17)$.

In addition, the role of inefficient B-to-D back transfer for $a$, $b$, and $c$ is reversed for the random distribution $d$. For $a$, $b$, and $c$, with a small $k_{BD}$ and a large $k_{BA}$, the directional energy flow of D → B → A is generated, whereas for $d$ the flow is highly disturbed because the excitation undergoes many trapping events at B sites during its migration. The extreme trapping event occurs when $k_{BD} = 0$ and a B site is surrounded only by D sites, though there is a small chance to escape via long-range transfer to nonadjacent B sites. This argument can be validated by making the back-transfer as efficient as other transfers. When $k_{BD} = 1$, the magnitude of $q$ becomes similar to the values for $a$ and $b$.

Like Figure 4a, it would be also interesting to find the optimal number of acceptors for the ABD systems shown in Figure 5. We found that $q$ shows the nonmonotonic behavior like Figure 4a as the number of acceptors per disk $N_A^{disk}$ increases from 1 and reaches the maximum around $N_A^{disk} = 3$ irrespective of spatial distributions and values of $k_{BD}$ (data not shown). $N_A^{disk} = 3$ was used in ref 10. For the random distribution $d$, the optimal $N_A$ is equal to around 7 for $k_{BD} = 1$ and equal to around 11 for $k_{BD} = 0$ (the change rates of $q$ are very slight around these numbers). For $k_{BD} = 0$ at which the excitation migration is so sticky, it is expectable that more acceptors is helpful for the excitation to reach acceptors compared to the case of $k_{BD} = 1$.

Lastly, as shown in Figure 6, the ABD system is more efficient than the AD system. For $c$ and $d$, the numerical values of efficiency between $k_{BD} = 0$ and 1 are comparable to the experimental ones found in ref 10.

## V. Conclusions

The efficiency of energy transfer systems was examined by investigating the dependency on the system properties, including network connectivity, transfer rates, relative ratio of the components, and their spatial arrangement. The main results are summarized below: (1) We found the optimal efficiency through the introduction of photon capturing ability, which is expressed as a fraction of bright chromophores. For the optimal design of light-harvesting devices, the photon capturing ability is as important as the quantum yield defined by $P_A$. (2) The increase in efficiency with $k_{DA}$ is strongly modulated by the spontaneous decay rate. The slower the spontaneous decay, the steeper the increase in efficiency with $k_{DA}$. This feature can be explained by eq 5, which is equivalent to the formula given in ref 14. Such behavior implies that the introduction of bridge chromophores can be a useful method to improve the efficiency for AD systems given inefficient D-to-A transfer and slow spontaneous decay. (3) In the study of the stacked-disk rod geometry with spacings between sites fixed, we found the staggered conformation to be more efficient than the eclipsed conformation. (4) For a given ratio of components, the uniform distribution of acceptors to minimize the mean first passage time to acceptors is a key point to design the optimal network for an AD system with a relatively small $k_{DA}$. (5) For a three-component ABD system with a large $k_{BA}$, it is important to increase the number of the pathways in accordance with the directional energy flow from D to B to A chromophores.

Our analysis is motivated by the recent experiments performed in refs 10 and 11, and the results suggest that the reported synthetic light-harvesting system may well be the most efficient. Here we present a point-by-point comparison between the theoretical predictions and experimental facts: (1) The finding of an optimal ratio A/D for AD helical rods is in qualitative agreement with Figure 3e from ref 10. Further consideration of detailed structural data, anisotropy in energy transfer, and the experimental measure of efficiency may be needed to match the reported optimal ratio $N_D:N_A (= 33:1)$. (2) Figure 5a from ref 10 shows that the ABD disk rod with $N_D:N_B:N_A = 8:4:1$ is more efficient than the AD system with $N_D:N_A = 1:1$, which is comparable to Figure 6 of this paper. (3) Figure 4d suggests that the staggered conformation of the ABD disk rod in Figure 3e in ref 10 is more efficient than the eclipsed one. (4) Figure 3e in ref 10 lists values of the antenna effect, which measures the enhancement of acceptor emission due to energy transfer from donors. This enhancement is approximately proportional to the number of donors contributing to the emission of a single A, implying that for a given donor−acceptor ratio the uniform distribution of acceptors is favorable. This conclusion is consistent with the observations in Figures 4c and 6 for AD systems. (5) According to Figures 5 and 6, the spatial arrangement shown in Figure 5c in ref 10 is the best for positioning two AB disks. The ABD rod in Figure 5c can be a good candidate as a basic building block for constructing optimal light-harvesting devices. Because the present work is based on incoherent classical transfer kinetics, the effect of quantum coherence in the energy transfer networks will be investigated.

**Acknowledgment.** This work was supported by the Singapore-MIT Alliance for Research and Technology (SMART), the MIT Energy Initiative Seed Grant (MITEI), the MIT Exciton Center Seed Grant, and the National Science Foundation (0806266). We thank Young Shen for helping revise this submission.

## Appendix A: Derivations for the One-Dimensional AD System

The calculation of $P_A$ needs the explicit expression for the survival probability obtained from eq 1. In eq 1, $\mathbf{A}$ is diagonalized as $\mathbf{A} = \mathbf{Q} \cdot \mathbf{M} \cdot \mathbf{Q}^T$ with the eigenvector matrix $\mathbf{Q}$ and the eigenvalue matrix $\mathbf{M}$, which are given by[17]

$$(\mathbf{Q})_{ik} = \sqrt{\frac{2 - \delta_{k0}}{N + 1}} \cos\left[\frac{\left(i + \frac{1}{2}\right)k\pi}{N + 1}\right] \quad (A1)$$

$$(\mathbf{M})_{ik} = \delta_{ik}\mu_k = \delta_{ik}4\sin^2\left[\frac{k\pi}{2(N+1)}\right] \quad \text{(A2)}$$

The superscript T denotes the transpose. Assuming the initial condition that $p_i(0) = N_D^{-1}$, $\hat{S}(s)$ ($= \mathbf{u}^T \cdot \hat{\mathbf{P}}(s)$) is given as

$$\hat{S}(s) = [(s+k_S)\mathbf{1} + k_{DD}\mathbf{M} + k_{DA}\mathbf{W}]_{0,0}^{-1} \quad \text{(A3)}$$

where $\mathbf{1}$ denotes the unit matrix and $\mathbf{W} = \mathbf{Q}^T \cdot \mathbf{R} \cdot \mathbf{Q}$. Equation A3 can be expanded in terms of $k_{DA}$ as

$$\hat{S}(s) = (\mathbf{D}(s))^{-1} \cdot \sum_{n=0}^{\infty}[-k_{DA}\mathbf{W}\cdot\mathbf{D}(s)^{-1}]^n)_{0,0} \quad \text{(A4)}$$

where $\mathbf{D}(s) = (s+k_S)\mathbf{1} + k_{DD}\mathbf{M}$. The first few coefficients of eq A4 are explicitly given below:

$$(\mathbf{D}(s)^{-1})_{0,0} = \frac{1}{s+k_S} \quad \text{(A5a)}$$

$$(\mathbf{D}(s)^{-1}\cdot\mathbf{W}\cdot\mathbf{D}(s)^{-1})_{0,0} = \frac{1}{(s+k_S)^2}\frac{2}{N+1} \quad \text{(A5b)}$$

$$(\mathbf{D}(s)^{-1}\cdot[\mathbf{W}\cdot\mathbf{D}(s)^{-1}]^2)_{0,0} = \frac{1}{(s+k_S)^2}\frac{1}{N+1}\sum_{\alpha,\beta=0,N}C_{\alpha\beta}(s) \quad \text{(A5c)}$$

$$(\mathbf{D}(s)^{-1}\cdot[\mathbf{W}\cdot\mathbf{D}(s)^{-1}]^3)_{0,0} = \frac{1}{(s+k_S)^2}\frac{1}{N+1}\sum_{\alpha,\beta,\gamma=0,N}C_{\alpha\beta}(s)C_{\beta\gamma}(s) \quad \text{(A5d)}$$

where $C_{\alpha\beta}(s)$ is defined by

$$C_{\alpha\beta}(s) = \sum_{k=0}^{N}(\mathbf{Q})_{\alpha k}\frac{1}{s+k_S+\mu_k}(\mathbf{Q}^T)_{k\beta} \quad \text{(A6)}$$

Using the relations arising from the symmetry of the system that $C_{00}(s) = C_{NN}(s)$ and $C_{0N}(s) = C_{N0}(s)$ and the property that $(\mathbf{Q})_{Nk} = [(2-\delta_{k0})/(N+1)]^{1/2}(-1)^k\cos[k\pi/2(N+1)]$, the summation of $C_{\alpha\beta}(s)$ in the right-hand side of eq A5c can be rewritten as

$$\sum_{\alpha,\beta=0,N}C_{\alpha\beta}(s) = 2\frac{2}{N+1}\left[\frac{1}{s+k_S}+\hat{\chi}_1(s+k_S)\right] \quad \text{(A7)}$$

where $\hat{\chi}_1(s)$ denotes the Laplace transformation of $\chi_1(t)$. $\chi_1(t)$ is the pure relaxation part of the normalized sink−sink time correlation function[18,19] and is here calculated as

$$\hat{\chi}_1(s) = \sum_{k \text{ even}}\frac{2\cos^2[k\pi/2(N+1)]}{s+4k_{DD}\sin^2[k\pi/2(N+1)]} \quad \text{(A8)}$$

Taking into account that $C_{00}(s) = C_{NN}(s)$ and $C_{0N}(s) = C_{N0}(s)$, one can show that the general expressions of higher-order coefficients ($n \geq 1$) are given in a factorized form by

$$(\mathbf{D}(s)^{-1}\cdot[\mathbf{W}\cdot\mathbf{D}(s)^{-1}]^n)_{0,0} = \frac{1}{(s+k_S)^2}\frac{2}{N+1} \times \left(\frac{2}{N+1}\left[\frac{1}{s+k_S}+\hat{\chi}_1(s+k_S)\right]\right)^{n-1} \quad \text{(A9)}$$

which in fact corresponds to the Wilemski−Fixman closure approximation.[18−20] Using eq A9, eq A4 can be easily recast as

$$\hat{S}(s) = \frac{1}{s+k_S} - \frac{1}{(s+k_S)^2}\frac{2k_{DA}}{N+1} \times \left[1 + \frac{2k_{DA}}{N+1}\left(\frac{1}{s+k_S}+\hat{\chi}_1(s+k_S)\right)\right]^{-1} \quad \text{(A10)}$$

which reads after some arrangements

$$\hat{S}(s) = \left[s + k_S + \left[\frac{N+1}{2k_{DA}} + \sum_{k \text{ even}}\frac{2\cos^2[k\pi/2(N+1)]}{s+k_S+4k_{DD}\sin^2[k\pi/2(N+1)]}\right]^{-1}\right]^{-1} \quad \text{(A11)}$$

$P_A$ is then calculated with eqs 3 and A11. The large $N$ expression of eq A11 can have a much simpler form using the asymptotic expressions $\cos^2[k\pi/2(N+1)] \sim 1$ and $\sin^2[k\pi/2(N+1)] \sim (k\pi/2N)^2$ for $N \gg 1$ and the equality that $\sum_{k=1}^{\infty}(a+bk^2)^{-1} = (y\coth y - 1)/2a$ with $y = \pi(a/b)^{1/2}$. For this case, eq 4 reduces to

$$q_{\infty} = \left[\frac{N+1}{2k_{DA}/k_S} + y\coth(y)\right]^{-1}X_D \quad \text{(A12)}$$

where $y = (N/2)(k_S/k_{DD})^{1/2}$. The compact form of eq A12 enables us to find $N^*$ at which eq A12 reaches the maximum as a function of dimensionless scaled parameters $\tilde{k}_{DA}$ ($= k_{DA}/k_{DD}$) and $\tilde{k}_S$ ($= k_S/k_{DD}$). Differentiating eq A12 with respect to $N$, expanding the result in terms of $\tilde{k}_S$, keeping the series up to the first-order term in $\tilde{k}_S$, and equating the truncated series to zero, we obtain the cubic equation

$$N^3 + (5 + 3\tilde{k}_S)N^2 + (3 + 18\tilde{k}_{DA}^{-1})N - 12\tilde{k}_S^{-1} + 15\tilde{k}_{DA}^{-1} = 0 \quad \text{(A13)}$$

Taking the real solution of eq A13, expanding the solution in terms of $\tilde{k}_S$, and keeping the series up to the first-order term in $\tilde{k}_S$ as consistent to the previous truncation, we obtain the approximate expression of $N^*$ as

$$N^* = \left(\frac{12}{\tilde{k}_S}\right)^{1/3} - \frac{5}{3} - 18^{1/3}\left(\frac{1}{\tilde{k}_{DA}} - \frac{8}{27}\right)\tilde{k}_S^{1/3} + \frac{5}{12^{2/3}}\left(\frac{1}{\tilde{k}_{DA}} - \frac{23}{81}\right)\tilde{k}_S^{2/3} - \tilde{k}_S \quad \text{(A14)}$$

The approximate valid range of eq A14 is given as $\tilde{k}_{DA} \geq 0.1$ and $\tilde{k}_S \leq 0.01$. Equation A14 shows that $N^*$ increases with $\tilde{k}_S^{-1}$ and $\tilde{k}_{DA}$ and is significantly affected by $\tilde{k}_{DA}$ for small $\tilde{k}_{DA}$ but becomes independent of $\tilde{k}_{DA}$ in the large $\tilde{k}_{DA}$ limit. In the right-hand side of eq A14, note that the magnitude of the one-third-order term is larger than that of the two-third-order term for $\tilde{k}_S < 1$ and the signs of the two terms are reversed as $\tilde{k}_{DA}$ increases.

### Appendix B: Alternative Derivation of the Limiting Expression of Eq A12

The infinite-$k_{DA}$ limit of eq A12 can alternatively be obtained by using the Green's function of the one-dimensional diffusion equation with absorbing boundary conditions at both ends ($0 \leq x \leq L$), which is given by[21]

$$G(x,t|x_0) = \frac{2}{L}\sum_{n=1}^{\infty} \sin\frac{n\pi x}{L} \sin\frac{n\pi x_0}{L} e^{-\pi^2 n^2 D_1 t/L^2} \quad (B1)$$

with $D_1$ denoting the one-dimensional diffusion coefficient. With the lattice spacing $a$, $D_1/a^2$ and $L/a$ correspond to $k_{DD}$ and $N$, respectively.

Let $p(x,t)$ denote the one-time probability distribution function that a particle is found at $x$ and at time $t$. $p(x,t)$ is obtained by averaging eq B1 over the initial equilibrium distribution and the corresponding survival probability $S(t)$ is then obtained by integrating $p(x,t)$ over the whole range. The explicit expression of $S(t)$ is given by

$$S(t) = \frac{1}{L}\int_0^L dx \int_0^L dx_0\, G(x,t|x_0) = \frac{8}{\pi^2}\sum_{k\,\text{odd}}\frac{1}{k^2}e^{-\pi^2 k^2 D_1 t/L^2} \quad (B2)$$

which has the same form as the normalized time correlation function of the end-to-end vector of a Rouse chain.[17] Under the spontaneous decay occurring uniformly over the whole range of $x$, the corresponding survival probability is simply obtained by multiplying eq B2 by $e^{-k_S t}$. In such a case, the mean reaction time $\hat{S}(0)$ is given by

$$\hat{S}(0) = \frac{8}{\pi^2}\sum_{k\,\text{odd}}\frac{1}{k^2}\frac{1}{k_S + \pi^2 k^2 D_1/L^2} \quad (B3)$$

which can be rewritten using the equality that $\sum_{k\,\text{odd}}^{\infty}(ak^2 + bk^4)^{-1} = (\pi^2/8a)(1 - \tanh y/y)$ with $y = (\pi/2)(a/b)^{1/2}$ as

$$\hat{S}(0) = k_S^{-1}(1 - \tanh y/y) \quad (B4)$$

where $y = (N/2)(k_S/k_{DD})^{1/2}$. Substituting eq B4 into eq 3 yields the infinite-$k_{DA}$ limit of eq A12.

### Appendix C: Derivations for the One-Dimensional ABD System

The master equation corresponding to the one-dimensional ABD system given in Figure 1b is easily obtained by expanding the dimensions of the matrices in eq 1. Its explicit form is not given here but its solution is directly given below:

$$\hat{\mathbf{P}}(s) = \mathbf{Q}'\cdot[s + \mathbf{K}_S + k_{DD}\mathbf{M}' + \mathbf{W}']^{-1}\cdot\mathbf{Q}'^T\cdot\mathbf{P}(0) \quad (C1)$$

where $\hat{\mathbf{P}}(s)$ here denotes the $(N+3)$-dimensional column vector where the $n$th element ($0 \leq n \leq N+2$) is the Laplace-transformed probability that an excitation is located at the $n$th site at time $t$ (see Figure 2b). The initial vector $\mathbf{P}(0)$ is given as $p_n(0) = (N_D + 2)^{-1}$ for bright B and $p_n(0) = N_D^{-1}(1 - \delta_{0,n} - \delta_{N+2,n})$ for dark B. $\mathbf{K}_S$ and $\mathbf{M}'$ are the $(N+3) \times (N+3)$ diagonal matrices whose diagonal elements are given by $(k_{SB}, k_{SD}, k_{SD}, ..., k_{SD}, k_{SB})$ and $(0, \mu_0, \mu_1, ..., \mu_N, 0)$, respectively. $k_{SB}$ and $k_{SD}$ denote the spontaneous relaxation rates of B and D, respectively. $\mathbf{W}' = \mathbf{Q}'^T\cdot\mathbf{R}'\cdot\mathbf{Q}'$, where

$$\mathbf{Q}' = \begin{pmatrix} 1 & \mathbf{0}^T & 0 \\ \mathbf{0} & \mathbf{Q} & \mathbf{0} \\ 0 & \mathbf{0}^T & 1 \end{pmatrix} \quad (C2)$$

and

$$\mathbf{R}' = \begin{pmatrix} k_{BA} + k_{BD} & -k_{DB} & \mathbf{0}^T & 0 & 0 \\ -k_{BD} & k_{DB} & \mathbf{0}^T & 0 & 0 \\ \mathbf{0} & \mathbf{0} & \mathbf{0}_M & \mathbf{0} & \mathbf{0} \\ 0 & 0 & \mathbf{0}^T & k_{DB} & -k_{BD} \\ 0 & 0 & \mathbf{0}^T & -k_{DB} & k_{BA} + k_{BD} \end{pmatrix} \quad (C3)$$

$\mathbf{0}$ and $\mathbf{0}_M$ are the column vector and the square matrix employed for filling the zero-element blocks for the given matrices.

For the general case with $k_{SB} \neq k_{SD}$, the expression of $P_A$ corresponding to eq 3 has a little bit different form, which is given by

$$P_A = 1 - k_{SD}\sum_{n=1}^{N+1}\hat{p}_n(0) - k_{SB}\hat{p}_0(0) - k_{SB}\hat{p}_{N+2}(0) \quad (C4)$$

where $\hat{p}_0(s) = \hat{p}_{N+2}(s)$ because of the symmetry of the system. In the right-hand side of eq C1, $\mathbf{Q}'^T\cdot\mathbf{P}(0)$ is calculated as

$$\mathbf{Q}'^T\cdot\mathbf{P}(0) = \frac{1}{N_D + 2}(1, N_D^{1/2}, 0, ..., 0, 1) \quad \text{(for bright B)} \quad (C5a)$$

$$\mathbf{Q}'^T\cdot\mathbf{P}(0) = \frac{1}{N_D}(0, N_D^{1/2}, 0, ..., 0, 0) \quad \text{(for dark B)} \quad (C5b)$$

Using eq C5 and $\sum_{n=1}^{N+1}(\mathbf{Q}')_{nk} = N_D^{1/2}\delta_{1k}$, eq C4 reduces to

$$P_A = 1 - k_{SD}\left[\frac{N_D}{N_D + 2}\mathbf{G}_{1,1} + \frac{2\sqrt{N_D}}{N_D + 2}\mathbf{G}_{1,0}\right] -$$

$$2k_{SB}\left[\frac{1}{N_D + 2}\mathbf{G}_{0,0} + \frac{\sqrt{N_D}}{N_D + 2}\mathbf{G}_{0,1} + \right.$$

$$\left. \frac{1}{N_D + 2}\mathbf{G}_{0,N+2}\right] \quad \text{(bright B)} \quad (C6a)$$

$$P_A = 1 - k_{SD}\mathbf{G}_{1,1} - 2k_{SB}\frac{1}{\sqrt{N_D}}\mathbf{G}_{0,1} \quad \text{(dark B)}$$

(C6b)

where $\mathbf{G} = [\mathbf{K}_S + k_{DD}\mathbf{M}' + \mathbf{W}']^{-1}$. Using eq C6 with the corresponding fraction of bright species, we can calculate eq 4 for the ABD system.